\documentclass[10pt,twocolumn,letterpaper]{article}

\usepackage{cvm}
\usepackage{times}
\usepackage{epsfig}
\usepackage{graphicx}
\usepackage{amsmath}
\usepackage{amssymb}
\usepackage{booktabs} 
\usepackage{bbding}
\usepackage[table]{xcolor}

\usepackage[pagebackref=true,breaklinks=true,letterpaper=true,colorlinks,bookmarks=false]{hyperref}

\newcommand{\keywords}[1]{{\bf \emph{Keywords: #1}}}

\cvmfinalcopy

\def\cvmPaperID{****} 

\ifcvmfinal\fi
\begin{document}

\title{CTSN: Predicting Cloth Deformation for Skeleton-based Characters with a Two-stream Skinning Network}

\author{Yudi Li\\
Zhejiang University\\
Hangzhou, China\\
\and
Min Tang\\
Zhejiang University\\
Hangzhou, China\\
%
\and
Yun Yang\\
Zhejiang University\\
Hangzhou, China\\
%
\and
Ruofeng Tong\\
Zhejiang University\\
Hangzhou, China\\
%
\and
Shuangcai Yang\\
Tencent\\
Shenzhen, China\\
%
\and
Yao Li\\
Zhejiang University\\
Shenzhen, China\\
%
\and
Bailin An\\
Zhejiang University\\
Shenzhen, China\\
%
\and
Qilong Kou\\
Zhejiang University\\
Shenzhen, China\\
}

\maketitle

\begin{abstract}
    We present a novel learning method to predict the cloth deformation for skeleton-based characters with a two-stream network. The characters processed in our approach are not limited to humans, and can be other skeletal-based representations of non-human targets such as fish or pets. We use a novel network architecture which consists of skeleton-based and mesh-based residual networks to learn the coarse and wrinkle features as the overall residual from the template cloth mesh. Our network is used to predict the deformation for loose or tight-fitting clothing or dresses. We ensure that the memory footprint of our network is low, and thereby result in reduced storage and computational requirements. In practice, our prediction for a single cloth mesh for the skeleton-based character takes about $7$ milliseconds on an NVIDIA GeForce RTX 3090 GPU. Compared with prior methods, our network can generate fine deformation results with details and wrinkles.
\end{abstract}


\section{Introduction}

\begin{figure}
 \includegraphics[width=\linewidth]{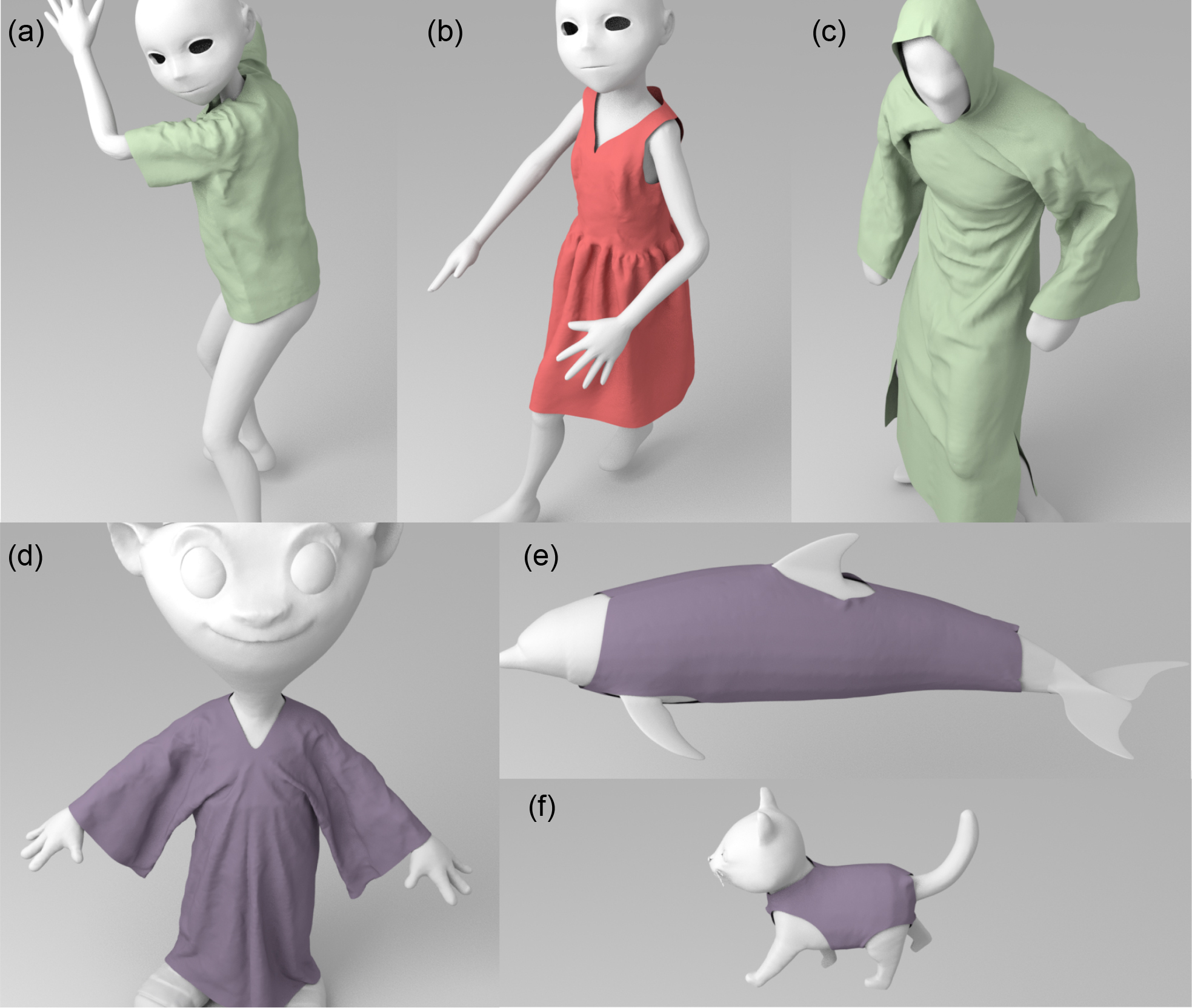}
 \centering
 \caption{Given a skeleton-based representation of a character corresponding to  target poses and different types of cloth (loose or tight-fitting),  we use a two-stream skinning network to predict the cloth deformation for the target character. (a) and (b) correspond to the same human character with tight and loose-fitting  clothing, respectively; (c) is a different human character wearing a long robe. Our network can also handle non-human characters such as a monster (d), a dolphin (e), or even a cat (f).}
 \label{fig:teaser}
\end{figure}

Cloth animation is an important problem in computer graphics due to its wide range of applications, including video games, special effects, and virtual try-on.  It is regarded  as a challenging task due to the model complexity of the cloth and the ability to perform irregular cloth deformations. Furthermore, many applications require interactive performance on commodity hardware, including mobile devices. This problem has been extensively studied in the literature. In order to achieve high-quality and reliable results, many efficient techniques based on physics-based simulation (PBS) have been proposed~\cite{baraff1998large, Provot95, Bouaziz2014, Bridson02, Harmon08, Tang14,tang2018cloth}. In these methods, the underlying cloth is modeled as a 3D surface mesh subdivided into finite contiguous triangles, and they use collision handling methods to generate accurate simulations. However, these methods cannot provide real-time frame rates for interactive applications.

There has been considerable work on using machine learning methods to significantly reduce the computational cost of predicting cloth deformation. Many learning-based networks~\cite{patel2020tailornet, bertiche2021deepsd, santesteban2019learning} have been proposed for SMPL-based parametric 3D human models~\cite{loper2015smpl}. These SMPL-based methods are used to generate smooth deformations for humans moving with tight-fitting clothes. The prediction is generated in real-time because of the small number of parameters used in  SMPL-based networks. However, the SMPL-based model is limited and cannot be used on arbitrary objects or characters used in games. In order to handle more general characters and enhance the quality of prediction, other algorithms use multi-layer perceptron (MLP) models on the vertices of the cloth mesh to learn the deformation~\cite{wang2019learning}. Without using the topologies of a cloth mesh, such MLP-based method tends to train a network with a large  number of parameters, which increases the  memory overhead  and the runtime cost.
Recently, Graph Convolutional Networks (GCN) have been used to predict the cloth draping results on the human characters ~\cite{gundogdu2019garnet, gundogdu2020garnet++, vidaurre2020fully}. In practice, these methods need the pre-deformed cloth for the target pose~\cite{gundogdu2019garnet, gundogdu2020garnet++} or can only process the draping results on human characters in a T-pose~\cite{vidaurre2020fully}.
 
In this paper, we deal with skeleton-based characters, which are widely used in computer games and other interactive applications. These include human-like characters (such as leading roles), monster characters similar to humans (such as trolls), and animal characters (such as pets). All these different characters can wear different types of clothes. We propose a learning-based cloth skinning model to capture the coarse and wrinkle features to obtain the final cloth deformation. Our approach is general and designed for all types of skeleton-based characters, including humans and animals. Furthermore, these characters can be dressed with loose or tight-fitting clothes.

Our formulation models the cloth draping deformation as the skinning of the cloth template at a canonical pose (such as a T-pose or a A-pose). Given the skeleton information and mesh information of the posed character, the deformation of the cloth is computed by skinning weights and the template cloth mesh. In order to handle different skinning characters and cloth, we design a novel two-stream network architecture to learn the residual positions of vertices of the cloth template mesh. It consists of a mesh-based residual stream and a skeleton-based residual stream. The skeleton-based residual stream is trained to obtain the coarse residual on the cloth template mesh, while the mesh-based residual stream is trained for the wrinkle features. The prediction examples of our two-stream skinning network are as show in ~\ref{fig:teaser}.

We qualitatively and quantitatively analyze the performance of the proposed two-stream skinning network in a variety of scenarios. These include human-like characters and other characters. We validate our two-stream network thorough the ablation experiments. Compared with recent methods, our two-stream network can capture the fine details of the cloth deformation.

The novel components of our work include:
\begin{itemize}
\item {\textbf{A learning-based cloth skinning model:} Our approach models the cloth deformation as the learning-based skinning of the template cloth mesh. Our skinning model is not limited to humans and can process many skinning characters.}
\item {\textbf{Two-stream skinning network architecture for cloth deformation prediction:} 
Based on the learning-based cloth skinning model, we design a novel two-stream network architecture for cloth deformation prediction. The architecture consists of a mesh-based residual stream which is trained for wrinkle features, and a skeleton-based residual stream which is trained for coarse features.}
\item {\textbf{Ability to process different types of clothes and characters:} Our network can process various types of characters and clothes. These characters and clothes can vary considerably.}
\end{itemize}

We show the prediction results of our proposed skinning-based network on different human characters, non-human characters with different cloth types in Section~\ref{sec:s5}. We compare our method qualitatively and quantitatively with other methods in Section~\ref{sec:s6}. We can predict deformed clothes at averagely $7$ milliseconds on an NVIDIA GeForce RTX 3090 GPU. As compared with prior approaches, our method can predict the deformation results with fine wrinkles and details.

\section{Related Work}
\label{sec:s2}

In this section, we give an overview of  cloth deformation prediction using traditional PBS methods and recent learning-based methods. Many learning-based methods are limited to the SMPL model; we describe these methods in Section~\ref{sec:s22} and highlight other learning methods in Section~\ref{sec:s23}.

\subsection{Physics-based Simulation}

PBS methods for generating deformed cloth are commonly based on the pipeline of time integration~\cite{baraff1998large}, collision detection~\cite{Bridson02, Tang14}, and collision response~\cite{Bridson02, Harmon08, tang2018cloth}. While they can accurately model the deformation and result in non-penetrating simulations, the running time is not fast enough for interactive applications.  To accelerate the simulation, recent research tends to use GPU-based algorithms to parallel the pipeline~\cite{tang2016cama, li2020p}. However, current methods can simulate each frame in hundreds of milliseconds on high-end desktop GPUs. Moreover, the performance of these simulators depends on various parameters, such as material attributes, which are hard to fine tune.

\subsection{SMPL-based Learning Algorithm}
\label{sec:s22}

Many learning methods have been proposed based on SMPL-based parametric 3D human models.
~\cite{loper2015smpl} proposed parametric skinning human models using SMPL, where the deformation of the human body mesh is driven by the skinning skeleton of the template body mesh. ~\cite{patel2020tailornet} regard the cloth mesh as the sub-mesh of the SMPL body mesh, and use an indicator matrix to select the associated vertices on the body mesh as the initial state. The proposed network, TailorNet~\cite{patel2020tailornet}, is trained as an increment from the initial state to represent the template cloth mesh. This is used to perform skinning operations to obtain the final deformation on the target pose.  \cite{li2021detail} use the skinning body mesh directly on the target pose as the initial state and learn a graph-attention-based network to predict the residual between the initial state and the final deformed cloth mesh with wrinkles.  These methods use the vertices on the unposed template body mesh or posed target body mesh as the initial state of the deformed cloth mesh and train different networks to fit the residuals of the ground truth. Therefore, the predictions of these methods may not generate plausible results on some loose-fitting clothes such as dresses, because the vertices may be far away from the body mesh.

Other algorithms have been proposed that treat the cloth mesh deformation as a skinning deformation similar to the body mesh skinning~\cite{Kavan07, loper2015smpl}. These methods tend to build a skinning model for cloth deformation from the canonical template cloth mesh. ~\cite{santesteban2019learning} use a garment fit regressor and a garment wrinkle regressor to learn the nonlinear residuals of the ground truth from the canonical cloth mesh. To enhance the performance on loose-fitting clothes, ~\cite{santesteban2021self} smoothly diffuse the skinning parameters of neighbors for each vertex on the unposed cloth mesh. They propose an optimization-based strategy to project ground-truth garments to the canonical space without introducing collisions. However, the diffusion of the skinning parameters is only operated on the unposed canonical cloth, which makes the improvement of the predictions on the loose-fitting clothes limited. ~\cite{bertiche2021deepsd} use GCN to extract features on the unposed canonical cloth mesh to learn the blend weights. These methods ignore the impact of the poses on the skinning weight parameters. In practice, all these networks are constrained by the pose and shape parameters of SMPL.

\subsection{Learning-based Cloth Deformation}
\label{sec:s23}
Many learning-based methods have been proposed for general cloth meshes and characters that are not limited to SMPL-based representations. ~\cite{gundogdu2019garnet, gundogdu2020garnet++} use dual quaternion skinning (DQS)~\cite{Kavan07} to generate the pre-deformation of the cloth template from the canonical pose and use GCN blocks to learn the residuals from the pre-deformation to the ground truth cloth mesh. ~\cite{holden2019subspace} use the PCA to obtain the subspace of the cloth and the obstacle and use MLP to regress the non-linearity in  subspace deformation. Unfortunately, using the previous predictions as the input of the subsequent predictions  will accumulate the error and hinder the quality of the result. ~\cite{wang2019learning} only use the vertex coordinate of the cloth mesh to learn a cloth descriptor that can be fused with motion in latent space. Considering the difficulty of predicting the cloth deformation caused by body pose, ~\cite{vidaurre2020fully} use an encoder and decoder architecture with GCN to learn the draping effect of different cloth types on the canonical pose. Other methods are designed for general triangle mesh-based obstacles~\cite{holden2019subspace,li2021n}.
  
Many techniques have been proposed to estimate a collision-free subspace of general 3D deformable models and used to compute collision-free cloth configurations~\cite{tan2020lcollision,tan2021active}. For human-like characters, many learning methods~\cite{gundogdu2019garnet,bertiche2021pbns} use collision loss to penalize penetrated garment-body pairs during training.  Our approach for handling arbitrary characters and clothing types is complimentary and can be combined with these methods.

\section{CTSN: Our Approach}
\label{sec:s31}
\begin{figure*}[h]
  \centering
  \includegraphics[width=\linewidth]{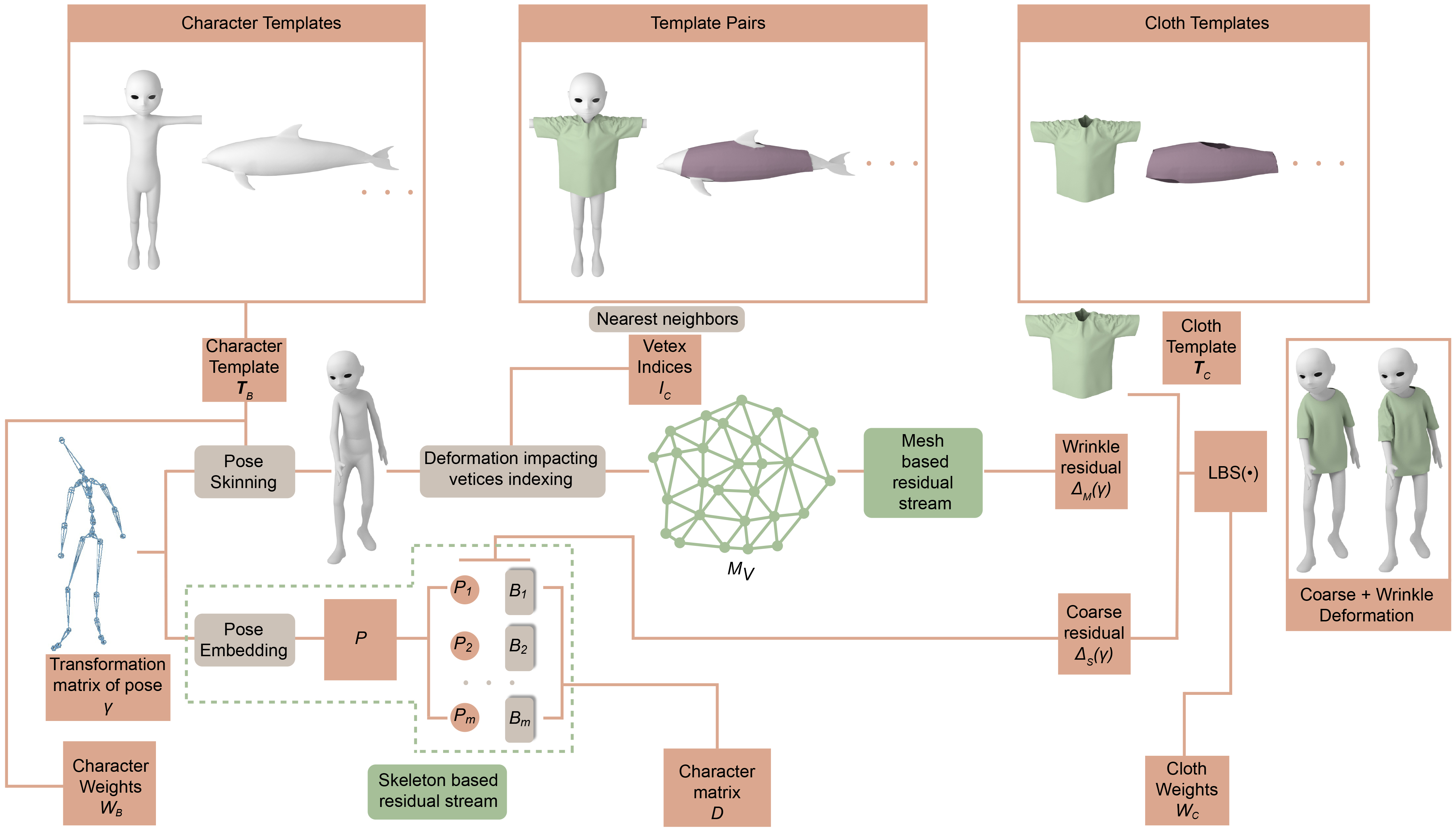}
  \caption{\label{fig:arch}
  Our network architecture is composed of the mesh-based residual stream and the skeleton-based residual stream (shown as the green blocks) to obtain the wrinkle residual $\Delta_{M}(\gamma)$ and the coarse residual $\Delta_{S}(\gamma)$. $\gamma$ is the transformation matrix of the target pose. The updated cloth template mesh $T_{C}(\gamma)$ is used by the skinning operation to obtain the final deformed cloth mesh $M_{C}(\gamma)$.
  }
\end{figure*}

Our approach takes a skeleton-based character of the target pose and cloth template of the canonical pose as input and predicts cloth mesh deformation for the target pose character through a skinning-based network. The skeleton-based character of the target pose has the skinned mesh and the transformation information of the joints. The key concept of our approach is a novel skinning-based cloth model. We propose a network architecture composed of two residual networks based the cloth model. We present the details of our skinning-based cloth model and the network architecture in following sections. 

\subsection{Skinning-based Cloth Model}

\subsubsection{Skinning-based Character Model}

Our skinning-based cloth model is inspired by the skinning-based character model, SMPL~\cite{loper2015smpl}. We give a brief overview of the SMPL model and the symbols used in the rest of the paper.

In the standard skeletal rigging, the posed character is calculated by the following formula:
\begin{equation}\label{}
M_{B}(\gamma) =\mathbf{W}\left(T_{B}, J, \gamma, W_{B}\right)
\end{equation}
where $M_{B}(\gamma)$ is the posed character mesh; $T_{B}$ is the template character mesh at the canonical pose; $J$ is the skeleton of character; $\gamma$ is the transformation matrix of the character joints; $W_{B}$ is the skinning weight matrix; and $\mathbf{W}(\cdot)$ is the skinning function. The parametric skinning human model SMPL~\cite{loper2015smpl} uses a set of orthonormal principal components of shape and pose displacements to capture the soft-tissue dynamics. This model is represented as:
\begin{equation}\label{}
\begin{aligned}
M_{B}(\beta, \theta) &=\mathbf{W}\left(T_{B}(\beta, \theta), J(\beta), \theta, W_{B}\right) \\
T_{B}(\beta, \theta) &=\mathbf{T}_{B}+B_{S}(\beta)+B_{P}(\theta)
\end{aligned}
\end{equation}
where $\beta$ and $\theta$ are the shape coefficients and the pose vector, which contains the transformation information of the joints, respectively. $J(\beta)$ is the skeleton position with shape coefficients $\beta$. $T_{B}(\beta, \theta)$ is the template human mesh, which is the function of $\beta$ and $\theta$. To capture the soft-tissue dynamics, body shape blend offsets $B_{S}(\beta)$ and pose blend shapes $B_{P}(\theta)$ are fused to the initial template human body mesh $\mathbf{T}_{B}$ to generate the final template human mesh $T_{B}(\beta, \theta)$.

\subsubsection{Our Two-stream Skinning-based Cloth Model}

Cloth deformation is driven by the character motion since cloth is dressed on the surface of a character mesh. To simplify the deformation problem, we use a skinning-based model for the template cloth mesh to guide the deformation. Inspired by the SMPL model and other approaches~\cite{santesteban2019learning}, we present a new method to build a skinning based model for cloth deformation. Thus, given a template cloth mesh $\mathbf{T_{C}}$ at the canonical pose and the skeleton transformation matrix at the target pose$\gamma$, the deformed cloth $M_{C}(\gamma)$ is defined as follows:
\begin{equation}\label{}
\begin{aligned}
M_{C}(\gamma) &=\mathbf{W}\left(T_{C}(\gamma), J, \gamma, W_{C}\right), \\
T_{C}(\gamma) &=\mathbf{T_{C}}+\Delta_{S}(\gamma)+\Delta_{M}(\gamma),
\end{aligned}
\end{equation}
where $\gamma$ is the transformation matrix of the joints of the target character body. $W_{C}$ is the skinning weight matrix for cloth template mesh $\mathbf{T_{C}}$. For the skinning function $\mathbf{W}(\cdot)$, $\operatorname{LBS(\cdot)}$ represents the linear blend skinning (LBS) method~\cite{Magnenat-thalmann88}, which is widely supported by game engines. $T_{C}(\gamma)$ is the optimized template cloth mesh at the canonical pose. $\Delta_{S}(\gamma)$ is the skeleton-based residual positions trained to obtain the coarse features. $\Delta_{M}(\gamma)$ is the mesh-based residual positions trained for adding wrinkle details to the coarse prediction. We highlight our two-stream network architecture in Fig.~\ref{fig:arch}. 

Our network architecture consists of a mesh-based residual stream and a skeleton-based residual stream. The mesh-based residual stream is designed to compute the impact of the nearest vertices of the cloth on the posed character mesh on the cloth template mesh, i.e. $\Delta_{M}(\gamma)$, while the skeleton-based residual stream is used to model the influence of skeleton information of the character to the cloth template mesh, i.e. $\Delta_{S}(\gamma)$. Since the cloth type can be tight or loose, we train the skinning weight matrix $W_{C}$ for different types of cloth. We present more details in Section~\ref{sec:s33}, ~\ref{sec:s34}, and ~\ref{sec:s35}.
In general, our network architecture can be expressed as:
\begin{equation}\label{eq:1}
M_{C}(\gamma) = \mathcal{N}_{\sigma}\left(\mathbf{T}_{C}, \mathbf{T}_{B}, J, \gamma, W_{B}, W_{C}\right),
\end{equation}
where $\mathcal{N}_{\sigma}$ is the skinning-based network and $\sigma$ represents the trainable parameters.

Similar to TailorNet~\cite{patel2020tailornet}, we decompose the deformed cloth mesh to the low-frequency and the high-frequency deformations. To obtain the low-frequency of the cloth mesh, we perform the Laplacian smoothing to the simulated cloth mesh. The high-frequency deformation is residual wrinkle details.

\subsection{Skeleton-based Residual Stream}
\label{sec:s33}

In our skeleton-based residual stream, the input is the transformation matrix $\gamma$ of character joints at the target pose. We pass the transform matrix $\gamma$ into the pose embedding network, which is composed of an MLP, to learn the pose embedding $\mathcal{P} = \{P_{1}, P_{2},  P_{2}, \cdots, P_{m}\}$, where $m$ is the size of the embedding vector $\mathcal{P}$:
\begin{equation}
\mathcal{P} = \Phi(\gamma),  \label{embedding}
\end{equation}
where $\Phi(\cdot)$ is the MLP-based pose embedding network. 

After the pose embedding, our goal is to learn a set of character residual matrices $D=\{B_{1}, B_{2}, B_{3}, \cdots B_{m}\}$ for the character and cloth pair.
As for matrix $B_{j}$, where $j \in \{1, 2, \cdots, m\}$, $B_{j}$ can be expressed as:
\begin{equation}
B_{j}=\left[\begin{array}{ccc}
b_{00} & \cdots & b_{02} \\
\vdots & \ddots & \vdots \\
b_{n0} & \cdots & b_{n2}
\end{array}\right],
\end{equation}
where $b_{00}, \cdots, b_{02}, \cdots, b_{n0}, \cdots, b_{n2}$ are trainable for the target character and cloth. $n$ is the number of vertices of the template cloth mesh.

Finally, the pose embedding $\mathcal{P}$ is fused as the weights to the residual matrix $D$ to obtain the skeleton-based residual component $\Delta_{S}(\gamma)$:
\begin{equation}
\Delta_{S}(\gamma)= \sum_{j=0}^{j=m} P_{j} B_{j}
\end{equation}

To train the skeleton-based residual stream to obtain the coarse features, we use the obtained low-frequency deformation as the ground truth.

\subsection{Mesh-based Residual Stream}
\label{sec:s34}

The skeleton-based residual stream can only predict the position offset $\Delta_{S}(\gamma)$, which captures the coarse features of the target deformation. The prediction results of the skeleton-based residual stream are smooth. To improve the prediction, we use a mesh-based residual stream to learn the wrinkle residual for the final cloth deformation.

We build a KD-tree for the template cloth mesh and the body mesh at canonical pose. We use this tree data structure to find the nearest point index $I_{C}$ on the body mesh for each vertex on the cloth mesh. Given the input transform matrix of the skeleton of the body, we can obtain the skinned body mesh at the target pose by using our skinning method. We obtain the positions $\mathcal{V}$ of the nearest points through the selected index $I_{C}$. In order to improve the effectiveness of our mesh-based residual stream, we also build the reference mesh graph $\mathcal{M_{V}}=(\mathcal{V}, \mathcal{E}, \mathcal{A})$, where $\mathcal{V}$ corresponds to the nearest vertices computed previously as the nodes of the graph $\mathcal{M_{V}}$; $\mathcal{E} \subseteq V \times V$ corresponds to the edges of the template cloth mesh, and $\mathcal{A}$ is the $(0, 1)$ adjacency matrix that highlights the connectivity of the vertices $\mathcal{V}$.

\begin{figure}[h]
  \centering
  \includegraphics[width=\linewidth]{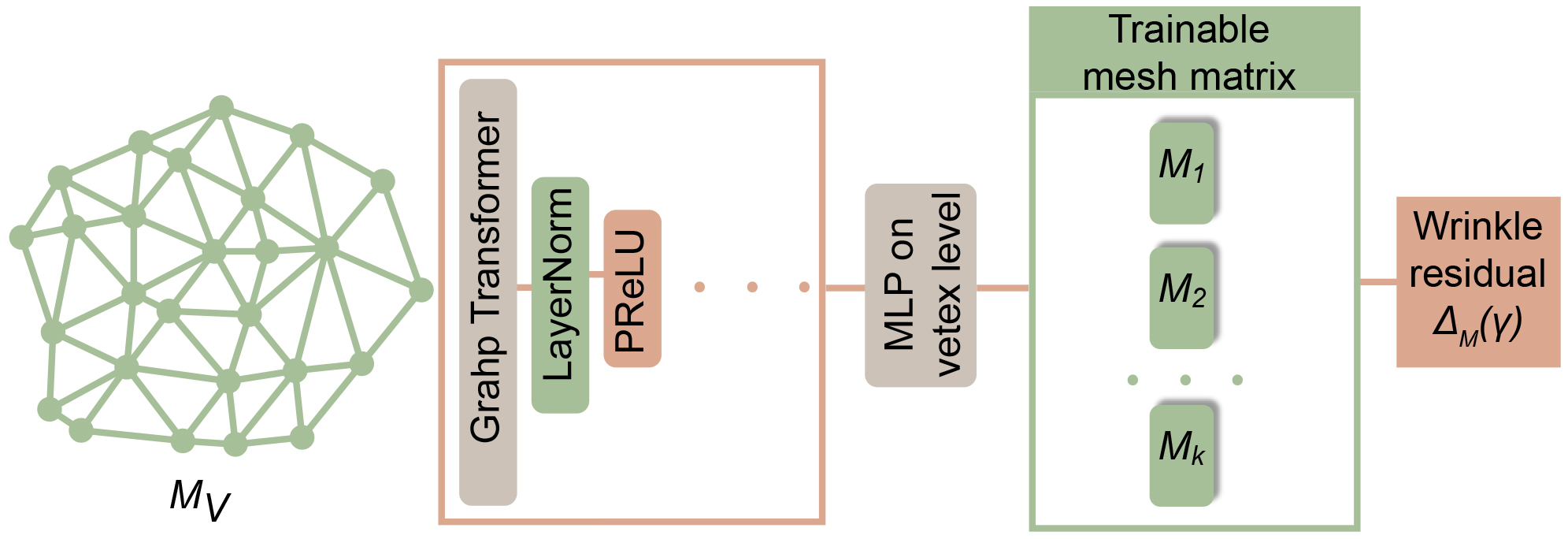}
  \caption{The architecture of our mesh-based residual stream. We use Transformer Graph Convolutional Network to extract features of the reference mesh graph $\mathcal{M_{V}}=(\mathcal{V}, \mathcal{E}, \mathcal{A})$.  The extracted features are transmitted to vertex level MLP layers and trainable mesh matrices to obtain the wrinkle residual.}
  \label{fig:transformer}
\end{figure}

We use the Graph Transformer network~\cite{shi2020masked} to extract features on the predefined constructed mesh graph $\mathcal{M_{V}}$. The architecture of the mesh-based residual stream is illustrated in Fig.~\ref{fig:transformer}. 

In the Graph Transformer layers of our mesh-based residual network, we define $H^{(l)}=\left\{h_{1}^{(l)}, h_{2}^{(l)}, \ldots, h_{n}^{(l)}\right\}$ as the node features of previous layer $l$, where $n$ is the number of nodes. $h_{i}^{l} \in \mathbb{R}^{F}$ represents the features of node $i$ in layer $l$ whose dimension is $F$. $h_{j}^{l}$ represents the features of node $j$ in layer $l$, where node $j$ is the neighbor of node $i$. The multi-head attention features $f_{c, i j}^{(l)}$ of head $c$ from node $j$ to node $i$ are computed as follows:
\begin{equation}
\begin{aligned}
q_{c, i}^{(l)} &= W_{c, q}^{(l)} h_{i}^{(l)}+b_{c, q}^{(l)} \\
k_{c, j}^{(l)} &= W_{c, k}^{(l)} h_{j}^{(l)}+b_{c, k}^{(l)} \\
e_{c, ij} &= W_{c, e} e_{i j}+b_{c, e} \\
f_{c, ij}^{(l)}&=(q_{c, i}^{(l)})^{\top}(k_{c, j}^{(l)} + e_{c, ij}) \\
\end{aligned}
\end{equation}
where $W_{c, q}^{(l)}$, $W_{c, k}^{(l)}$, $W_{c, e}$, $b_{c, q}^{(l)}$, $b_{c, k}^{(l)}$, and $b_{c, e}$ are trainable parameters. $e_{i j}$ represents the edge features. 

After normalization, the multi-head attention coefficients $\alpha_{c, i j}^{(l)}$ of head $c$ from node $j$ to node $i$ are computed as:
\begin{equation}
\begin{aligned}
\alpha_{c, i j}^{(l)}&=\operatorname{softmax}\left(\frac{f_{c, i j}^{(l)}}{\sqrt{d}}\right) \\
\end{aligned}
\end{equation}
where $d$ is the hidden size of each head.
The output features $\hat{h}^{(l+1)}$ of the node $i$ in layer $l + 1$ are calculated by the following formula:
\begin{equation}
\begin{aligned}
v_{c,j}^{(l)} &= W_{c, v}^{(l)} h_{j}^{(l)}+b_{c, v}^{(l)} \\
\hat{h}^{(l+1)} &= \|_{c=1}^{C}\left[\sum_{j \in \mathcal{N}(i)} a_{c, i j}^{(l)}\left(v_{c,v}^{(l)} + e_{c, ij}\right)\right] \\
\end{aligned}
\end{equation}
where $C$ is the number of the head. $W_{c, v}^{(l)}$ and $b_{c, v}^{(l)}$ are trainable parameters. $\mathcal{N}(i)$ is the neighbors of the node $i$. $\|$ is the concatenation operation for $C$ head attention.

In order to improve the ability of the feature extraction,
$\beta_{i}^{(l)}$ is calculated as follows:
\begin{equation}
\begin{aligned}
r_{i}^{(l)} &= W_{r}^{(l)} h_{i}^{(l)}+b_{r}^{(l)} \\
g_{i}^{(l)}&=W_{g}^{(l)}\left[\hat{h}_{i}^{(l+1)} ; r_{i}^{(l)} ; \hat{h}^{(l+1)}-r_{i}^{(l)}\right] \\
\beta_{i}^{(l)}&=\operatorname{sigmoid}\left(g_{i}^{(l)}\right) \\
\end{aligned}
\end{equation}
Thus, the final output features of the node $i$ in layer $l+1$ are updated as:
\begin{equation}
\begin{aligned}
r_{i}^{(l+1)}&=\left(1-\beta_{i}^{(l)}\right) \hat{h}_{i}^{(l+1)}+\beta_{i}^{(l)}\left(W_{r}^{(l)} h_{i}^{(l)}+b_{r}^{(l)}\right) \\
h_{i}^{(l+1)}&=\operatorname{ReLU}\left(\text { LayerNorm }\left(r_{i}^{(l+1)}\right)\right). \\
\end{aligned}
\end{equation}

As shown in Fig.~\ref{fig:transformer}, we use the Graph Transformer network to extract features of the mesh graph. After the feature extraction on the mesh graph, we use a vertex level MLP and a set of trainable mesh matrices to obtain the wrinkle residual positions. The trainable mesh matrices are represented as $\{M_{1}, M_{2}, M_{3}, \cdots M_{k}\}$.
$\operatorname{ReLU(\cdot)}$ is used to match the nonlinearity of the high-frequency deformation. $\Delta_{M}(\gamma)$ is computed from the mesh graph $\mathcal{M_{V}}$ as:
\begin{equation}
\Delta_{M}(\gamma)=\Psi(\mathcal{M_{V}}),
\end{equation}
where $\Psi(\cdot)$ represents the mesh-based residual stream. Similar to the skeleton-based residual stream, we use the high-frequency deformation as the ground truth to train the mesh-based residual stream.

\subsection{Skinning Operation}
\label{sec:s35}

After obtaining the skeleton-based residual component $\Delta_{S}$ and the mesh-based residual component $\Delta_{M}$, we compose a new optimized template cloth mesh $T_{C}(\gamma)$.

To solve the impact of cloth types (tight-fitting or loose-fitting)
on the final prediction results, we learn a weight residual $\Delta W_{C}$ for different cloth types. $\Delta W_{C}$ is represented as:
\begin{equation}
\Delta W_{C}=\left[\begin{array}{ccc}
w_{00} & \cdots & w_{0 k} \\
\vdots & \ddots & \vdots \\
w_{n_{0}} & \cdots & w_{n k}
\end{array}\right]
\end{equation}
where $w_{00} ,\cdots, w_{nk}$ are trainable parameters and $k$ is the maximum number of joints.

The fusion skinning weight matrix is generated as:
\begin{equation}
W_{C}= W_{C}^{I} + \Delta W_{C},
\end{equation}
where $W_{C}^{I}$ represents the initial cloth weight obtained from the template body skinning weight $W_{B}$ through KD-tree.

In general, pose embedding function $\Psi(\cdot)$ and $D$ are trained by skeleton-based residual stream for the coarse deformation, while $\Psi(\mathcal{M_{V}})$ is trained by mesh-based residual stream for the wrinkle deformation. $W_{C}$ is trained for processing different types of cloth.

\subsection{Loss Function}

To optimize the parameters of our network architecture, we use the following loss function to minimize the difference between the predicted deformed cloth mesh and the ground truth:
\begin{equation}\label{eq:loss}
\mathcal{L}=\frac{1}{\sum_{i = 1}^{b} N} \sum_{i = 1}^{b} \sum_{j=1}^{N}\left\|x_{p}^{j} - x_{g}^{j}\right\|_{2},
\end{equation}
where $x_{p}^{j}$ is the predicted position of vertex $j$ on the deformed cloth mesh $M_{CP}$. $x_{g}^{j}$ is the position of vertex $j$ on the ground truth cloth mesh $M_{CG}$. $N$ is the number of vertices of cloth mesh $M_{CG}$. $\left\|\cdots\right\|_{2}$ is the $L_{2}$ distance. $b$ is the batch size.

\section{Dataset and Implementation}
\label{sec:s4}

In this section, we describe the generation of our dataset and some implementation details.

\subsection{Dataset}

\begin{figure}[h]
  \centering
  \includegraphics[width=0.95\linewidth]{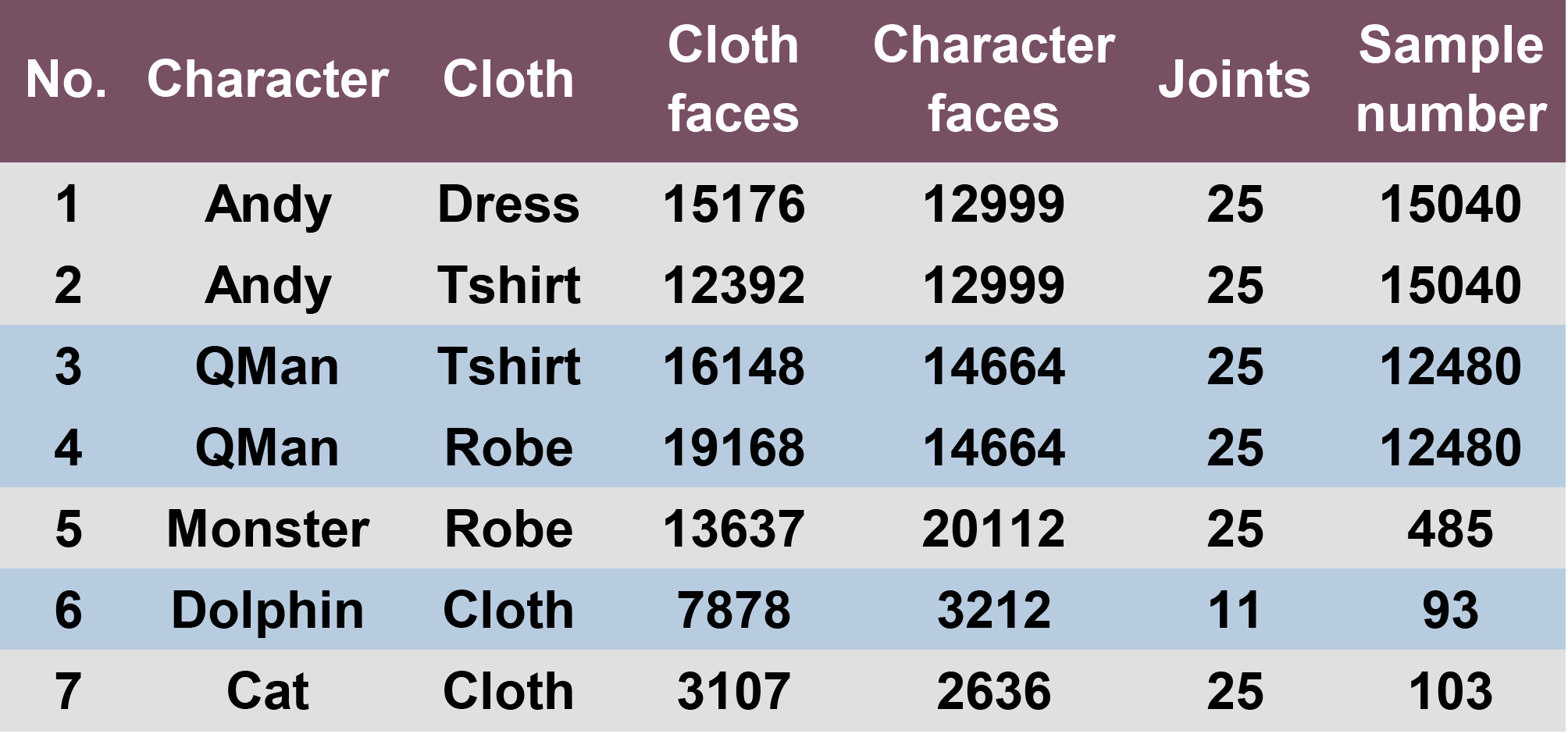}
  \caption{The attributes of different characters and clothing types used for our evaluation. We obtain different poses of characters from the Mixamo website. We extract the transformation matrix and skinning weight from the  motion files. We use the cloth simulator ARCSim
  to precompute the deformed cloth mesh for training.}
  \label{fig:dataset}
\end{figure}

We have generated many different characters and clothing types to validate our network architecture (as shown in Fig.~\ref{fig:dataset}). We upload the character meshes of  Andy and Qman in canonical poses, such as a T-pose, to the motion capture website Mixamo\footnote{https://www.mixamo.com/}. We download many character poses computed from that website as FBX files. To eliminate the absoluteness  of the vertex position and make it easy to train our network, we move the hip joint of the character mesh to the origin of the coordinates. Next, we extract the transformation matrix $\gamma$ of the character at different poses and the skinning weights $W_{B}$ from the FBX files.

After extracting of the motion files, we use the skinning operation to obtain the character meshes at different poses with transformation matrix of joint $\gamma$. We use different clothing types such as a T-shirt, dress, and robe. The T-shirt is tight-fitting, and the dress and robe are loose and can result in complex deformations.  In order to compute the ground truth of the deformed cloth, we use the physics-based simulator ArcSim~\cite{narain2012adaptive, narain2013folding, pfaff2014adaptive} to simulate the cloth. During the simulation, we perform linear interpolation between the adjacent poses and relax the cloth mesh to compute the quasi-static deformation.

To evaluate that our network can process more complex and different characters, we applied our network on non-human characters such as a monster, a dolphin, and a cat. The monster character has a skeleton similar to the human character, while the dolphin and the cat have different skeletons. The dolphin character has no leg joints, while the cat model has four legs without hands. We can also simulate the cloth deformation on these characters. The monster character wears a loose robe, and the dolphin and the cat wear tight-fitting clothes designed for these characters.

The attributes of the skinned character and cloth meshes are shown in Fig.~\ref{fig:dataset}. We have highlighted the number of triangles of each character mesh and cloth mesh, the number of joints of the character, and the number of samples used by our algorithm.

\subsection{Network Implementation and Training}

We train our network on a standard PC (Ubuntu 20.04 LTS/Intel I7 CPU@4.2G Hz/8G RAM, NVIDIA GeForce RTX 3090 GPU). Our network is implemented using PyTorch 1.7.0 and Python 3.8.8.

Following ~\cite{patel2020tailornet} and ~\cite{wang2019learning}, we also split our dataset for training and testing. For the motion clips obtained from Mixamo, we split $90\%$ motion clips as training data and the last $10\%$ motion clips as the test data, which are unseen during training.

We train our network on the dataset containing different characters and cloth types. As shown in Fig.~\ref{fig:dataset}, our dataset has 5 skeleton-based characters (2 human characters and 3 non-human characters) with 7 different types of cloth. During training, we set the learning rate at $1e-3$ and use an Adam optimizer~\cite{kingma2014adam} to train the parameters of the neural network.

\subsection{Penetration Handling}

It is hard to obtain collision-free predictions or configurations with learning-based methods on the test data, which is unseen during training. We use a method similar to ~\cite{wang2019learning} to reduce  the penetrations between the cloth and the character. After the prediction, the predicted deformed cloth mesh is optimized by minimizing the following function to avoid  penetrations between the cloth and the character:
\begin{equation}
E_{B}=\sum_{i \in {V_{pene}}}\left\|{v}_{i}-\left(v_{i}^{B}+\epsilon n_{i}^{B}\right)\right\|,
\end{equation}
where $V_{pene}$ is the set of penetrated vertices of predicted cloth. For each penetrated vertex $v_{i}$, the closest point vertex $v_{i}^{B}$ and normal $n_{i}^{B}$ are computed over the character mesh. $E_{B}$ is the error between penetration vertices on the cloth and the character mesh. and  $\epsilon$ is a small step to pull out the penetrated vertices from the character mesh. During the optimization process, the positions of $V_{pene}$ are updated, which reduces the number of localized penetrations or collisions.

\section{Results}
\label{sec:s5}

In this section, we highlight the deformation prediction results of our network on the unseen test data. We compare our predictions on the unseen test data with the ground truth results obtained using a physics-based simulator (ArcSim).

\subsection{Predicted Deformation using Our Network}

\begin{figure}[h]
  \centering
  \includegraphics[width=\linewidth]{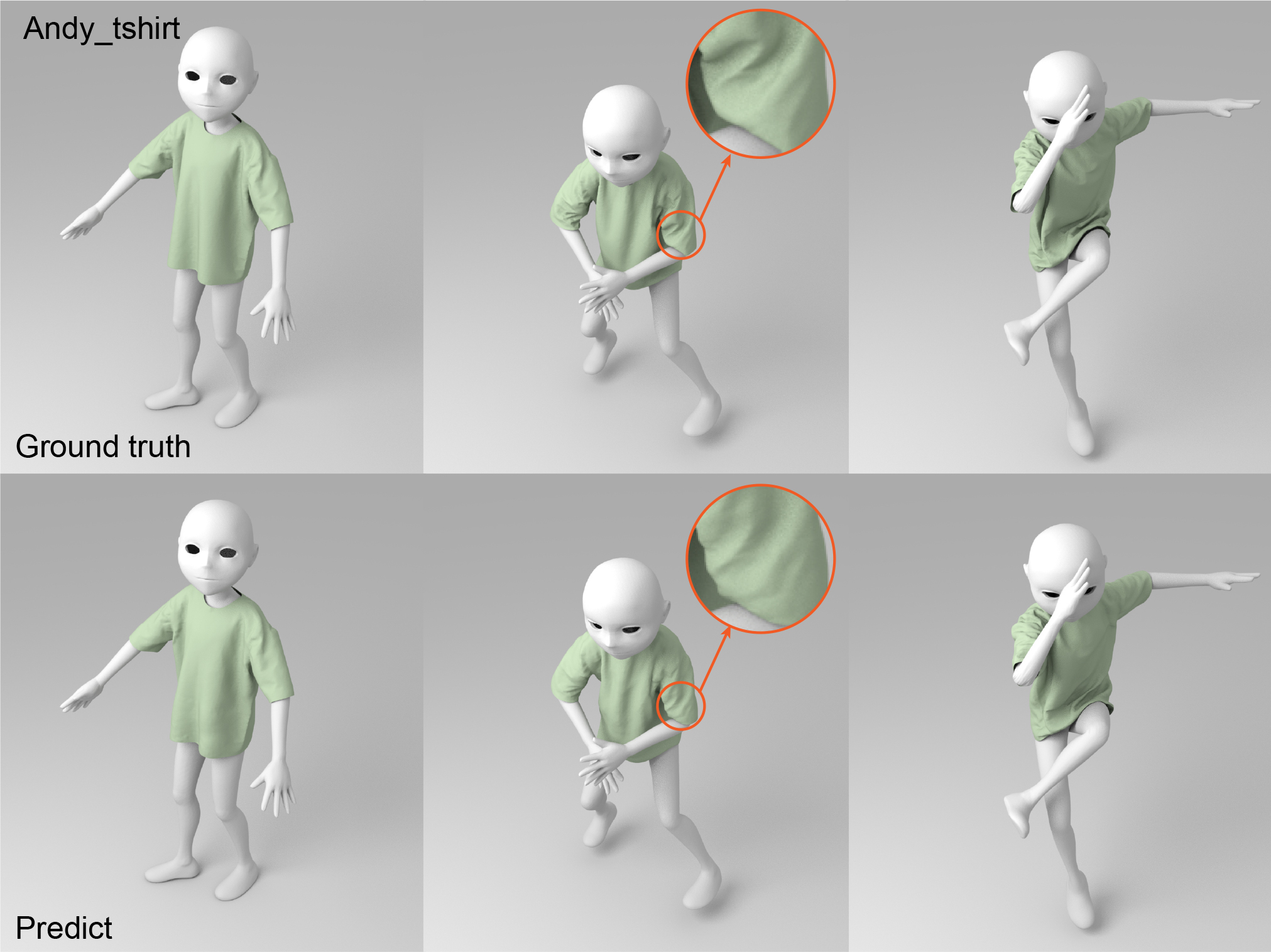}
  \caption{The predicted deformed T-shirt dressed on the character Andy in different poses. All the input poses are unseen during the network training. The top row shows the ground truth of the deformation, while the bottom row highlights the predictions of our network. We also highlight the fine details and folds in the zoomed images.}
  \label{fig:andy_re}
\end{figure}

\begin{figure}[h]
  \centering
  \includegraphics[width=\linewidth]{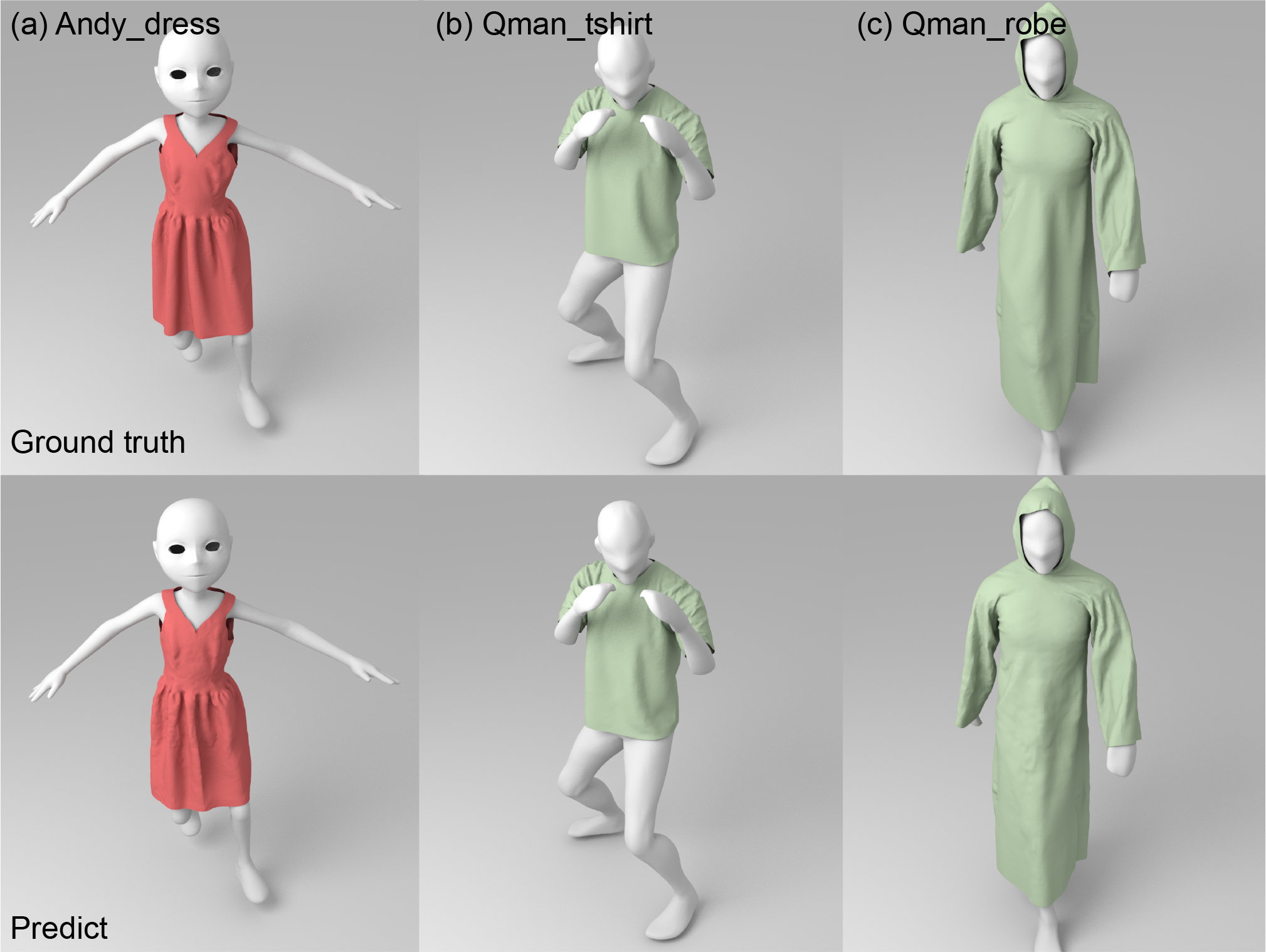}
  \caption{The predicted deformed cloth on other human characters. The first column shows the prediction on the character Andy. The middle and last columns show the deformation predictions on the character Qman.}
  \label{fig:andy_qman}
\end{figure}

Fig.~\ref{fig:andy_re} shows the predicted T-shirt deformation at different poses for the character Andy. Our predictions show the fine details with wrinkles, similar to those in the ground truth deformation. We also show the prediction results of other types of cloth and another character, Qman, in Fig.~\ref{fig:andy_qman}. Fig.~\ref{fig:andy_qman} (a) shows the predicted deformation of the dress on the character Andy, while Fig.~\ref{fig:andy_qman} (b) and (c) show the cloth deformation on the other character, Qman. The dress on the character Andy in Fig.~\ref{fig:andy_qman} (a) and the robe on the character Qman in Fig.~\ref{fig:andy_qman} (c) are both loose-fitting types of clothing. These predictions validate the effectiveness of our network.
Since we train the mesh-based residual stream and skinning weight for each clothing type, the deformation details can be easily captured, enhancing the predictions.

\begin{figure}[h]
  \centering
  \includegraphics[width=\linewidth]{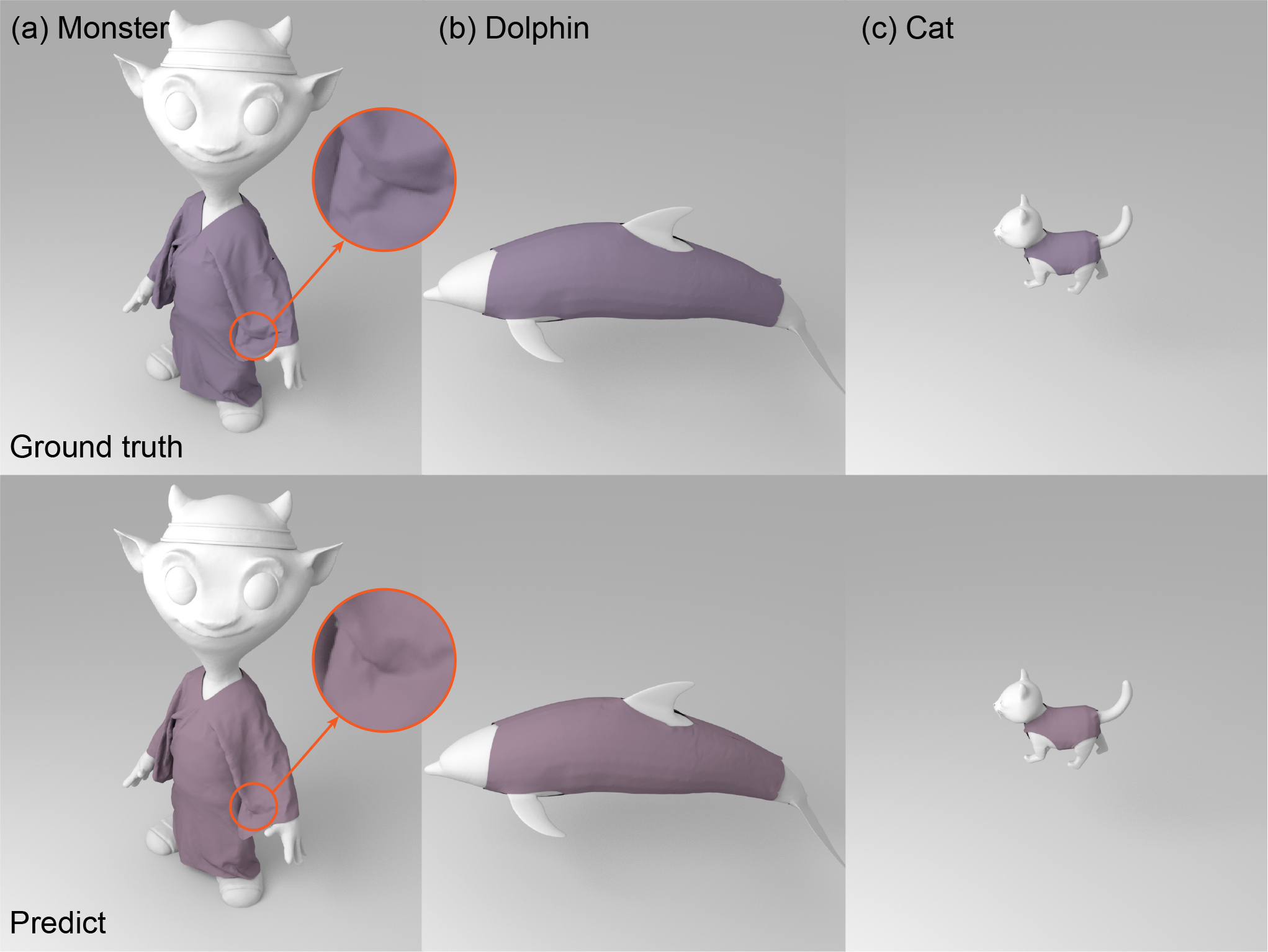}
  \caption{The results of our network on non-human characters.The first column shows the deformed robe on the Monster, whose skeleton is similar to that of human characters. The middle column shows the deformed cloth on the Dolphin, which has no legs. The last column shows the cloth on the Cat, which has no arms.}
  \label{fig:monster}
\end{figure}

\begin{figure}[h]
  \centering
  \includegraphics[width=\linewidth]{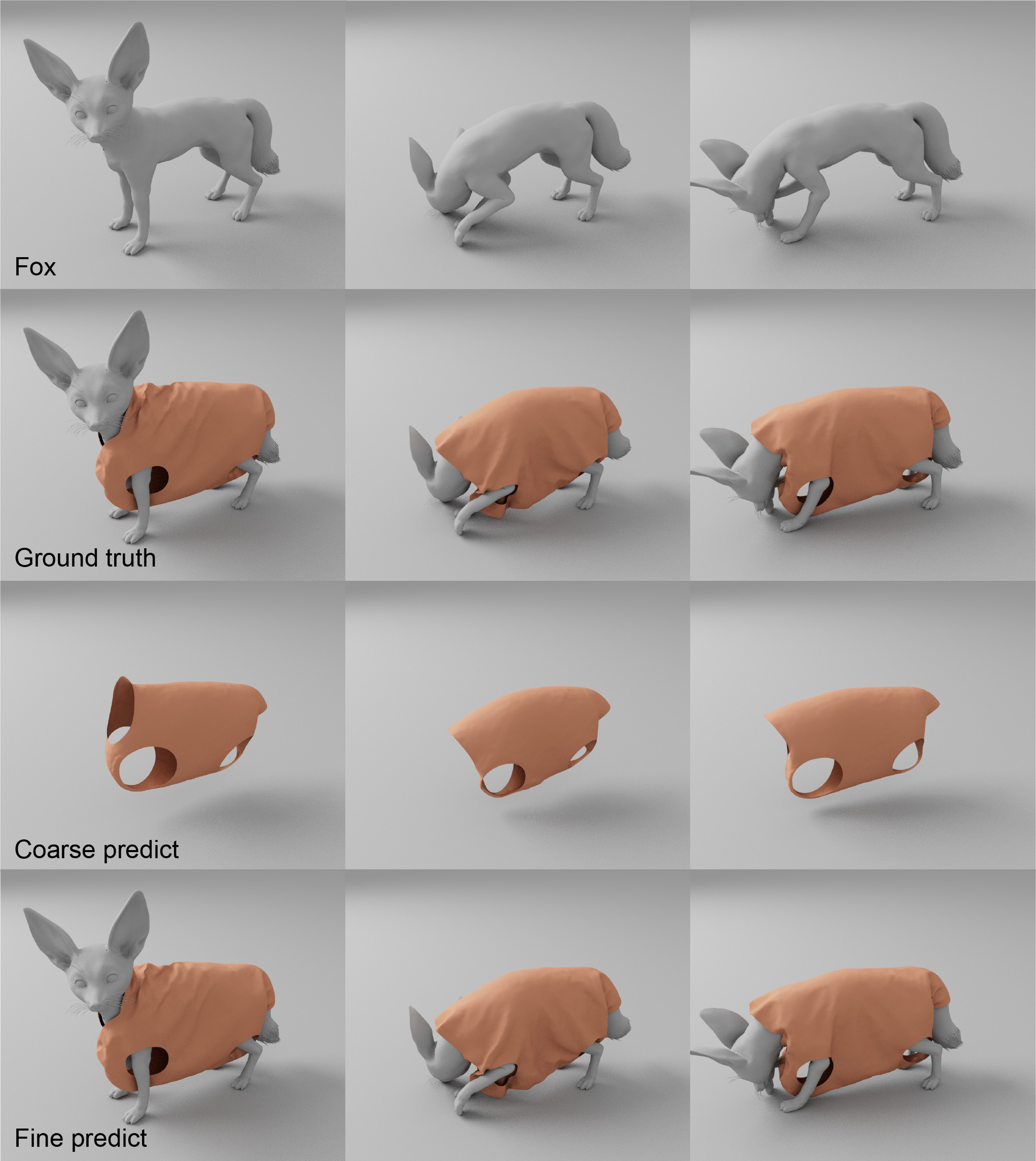}
  \caption{The results of our network on non-human character fox. There is a loose-fitting cloth dressed on the character fox. The first row shows the target pose of the character fox. The second row shows the ground truth. The third row shows the coarse prediction. The last row shows the fine detailed prediction.}
  \label{fig:fox}
\end{figure}

Our network can also process other non-human characters with skeletons. The predicted results of our network and the ground truth on the non-human characters are shown in Fig.~\ref{fig:monster}. Fig.~\ref{fig:monster} (a) shows the result of our network on a non-human character, Monster.  The skeleton hierarchy of Monster is similar to the human characters in Fig.~\ref{fig:andy_qman}. To show the complex characters that our network can process, we highlight the results of our network on the Dolphin character in Fig.~\ref{fig:monster} (b) and the Cat character in Fig.~\ref{fig:monster} (c). The Dolphin has no arm joints or legs joints, while the Cat has four legs without arms. The cloth on the Dolphin and the Cat are designed specifically for these characters. The results of our network show the fine predictions of the cloth deformations on these non-human characters. As for the non-human characters, the deformation of loose-fitting cloth is also well predicted. Fig.~\ref{fig:fox} shows the loose-fitting cloth dressed on the character Fox. Our network can predict the coarse deformed cloth and the fine detailed one. The results of cloth deformation on the Dolphin, the Cat and the Fox show the capability of our network processing non-human characters. The prediction of deformation can catch the fine wrinkle details.

\begin{figure}[h]
  \centering
  \includegraphics[width=\linewidth]{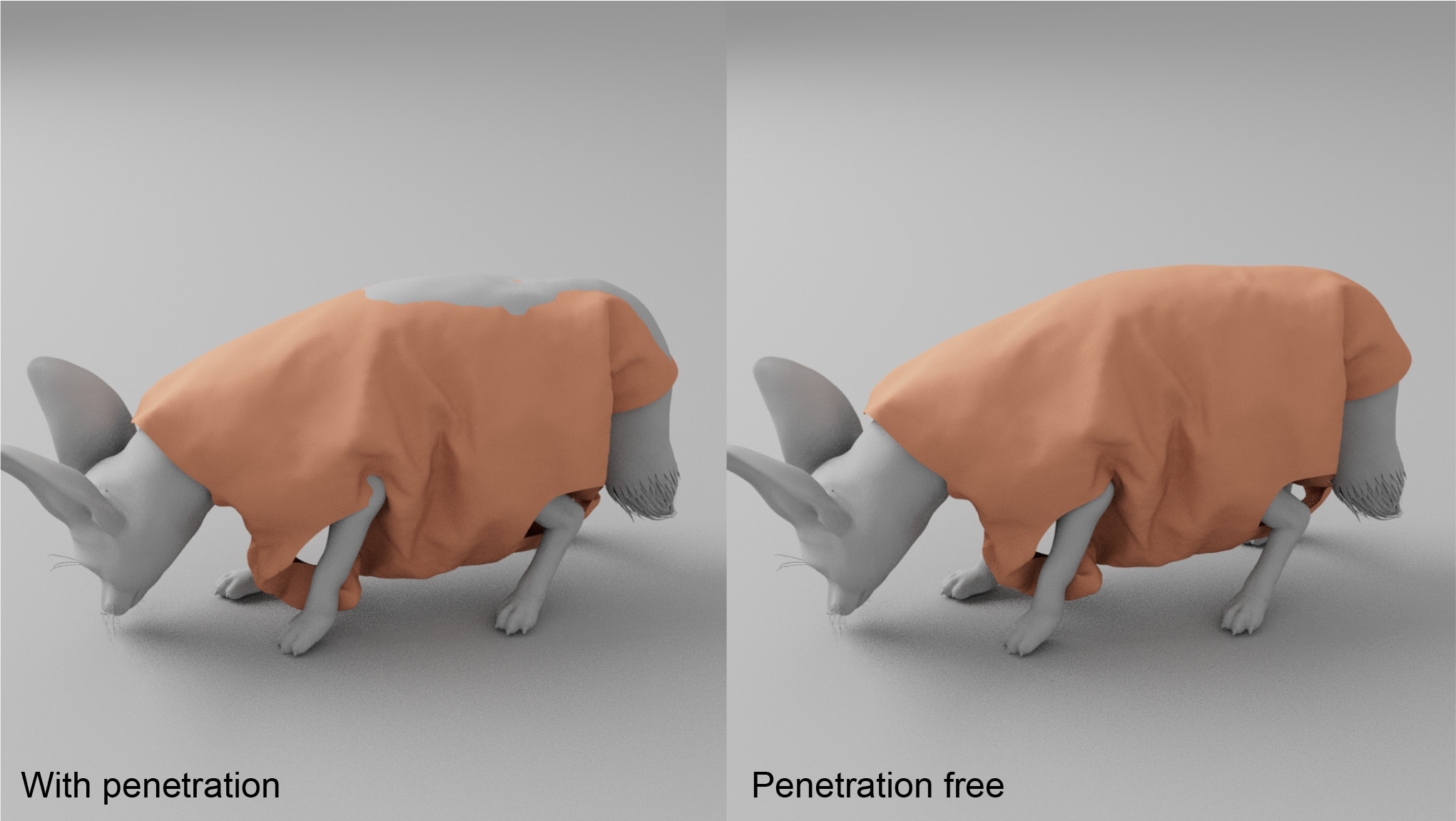}
  \caption{The results of penetration Handling. The left is the situation of penetration between the character fox and the loose-fitting cloth. The right is the penetration-free result.}
  \label{fig:pene}
\end{figure}

Fig.~\ref{fig:pene} shows the result of penetration handling described in Sec.4.3. After post-processing, the penetration between the back of the character fox and the dressed cloth is eliminated and the penetration-free result is obtained.

\subsection{Prediction Runtime}
We can perform cloth deformation prediction with our network both on GPUs and CPUs. We have highlighted the runtime of predicting a single cloth mesh in Table.~\ref{tab:time}. The runtime for a GPU is collected on an NVIDIA GeForce RTX 3090 GPU. The runtime for a CPU is collected on an Intel I7 CPU. As shown by the table, we can perform a single prediction within $7$ms on a GPU,
which is much faster than prior learning-based~\cite{li2021n} or physically-based algorithms~\cite{li2020p}. The running time of our deformation prediction algorithm on CPU in less than $0.2$s.


\begin{table}[h]
\begin{center}
\begin{tabular}{|l|c|c|}
\hline
Methods & CPU run time (s) & GPU run time (s) \\
\hline\hline
ARCSim & 3.45 & / \\
I-Cloth & / & 5.12E-2 \\
PBNS & 9.55E-2 & 7.12E-2 \\
DeePSD & 3.12E-1 & 1.25E-1 \\
Our Method & 1.72E-1  & 7.032E-3    \\
\hline
\end{tabular}
\end{center}
\caption{The average CPU and GPU runtime for a single cloth mesh prediction.}
\label{tab:time}
\end{table}

\section{Comparisons}
\label{sec:s6}

\begin{table*}
\begin{center}
\begin{tabular}{|l|c|c|c|c|c|c|c|}
\hline
Network for & SMPL & Non-SMPL & Non-human & Rigid & Static & Single \\
coth deformation & human & human & skinned animal & obstacle & prediction & network \\
\hline\hline
TailorNet\cite{patel2020tailornet} & \color{red}\CheckmarkBold & \color{red}\XSolidBrush &\color{red}\XSolidBrush &\color{red}\XSolidBrush & \color{red}\CheckmarkBold &\color{red}\XSolidBrush \\
    DeePSD \cite{bertiche2021deepsd} & \color{red}\CheckmarkBold & \color{red}\CheckmarkBold &\color{red}\CheckmarkBold &\color{red}\XSolidBrush & \color{red}\CheckmarkBold &\color{red}\XSolidBrush \\
    \cite{santesteban2019learning} & \color{red}\CheckmarkBold & \color{red}\XSolidBrush &\color{red}\XSolidBrush &\color{red}\XSolidBrush & \color{red}\XSolidBrush &\color{red}\XSolidBrush \\
    \cite{santesteban2021self} & \color{red}\CheckmarkBold & \color{red}\XSolidBrush &\color{red}\XSolidBrush &\color{red}\XSolidBrush & \color{red}\XSolidBrush &\color{red}\XSolidBrush  \\
    \cite{holden2019subspace} & \color{red}\CheckmarkBold & \color{red}\CheckmarkBold &\color{red}\CheckmarkBold &\color{red}\CheckmarkBold &\color{red}\XSolidBrush & \color{red}\XSolidBrush  \\
    GarNet \cite{gundogdu2019garnet} & \color{red}\CheckmarkBold & \color{red}\CheckmarkBold & \color{red}\XSolidBrush & \color{red}\XSolidBrush & \color{red}\CheckmarkBold & \color{red}\XSolidBrush  \\
    \cite{wang2019learning} & \color{red}\CheckmarkBold & \color{red}\CheckmarkBold & \color{red}\XSolidBrush & \color{red}\XSolidBrush & \color{red}\CheckmarkBold & \color{red}\XSolidBrush \\
    \cite{bertiche2021pbns} &\color{red}\CheckmarkBold & \color{red}\CheckmarkBold &\color{red}\CheckmarkBold &\color{red}\XSolidBrush & \color{red}\CheckmarkBold &\color{red}\XSolidBrush \\
    DRAPE \cite{guan2012drape} & \color{red}\CheckmarkBold & \color{red}\XSolidBrush &\color{red}\XSolidBrush &\color{red}\XSolidBrush &\color{red}\CheckmarkBold &\color{red}\XSolidBrush \\
    {N-Cloth \cite{li2021n}} & \color{red}\CheckmarkBold & \color{red}\CheckmarkBold & \color{red}\CheckmarkBold & \color{red}\CheckmarkBold & \color{red}\CheckmarkBold & \color{red}\XSolidBrush  \\
    {Our method} & \color{red}\CheckmarkBold & \color{red}\CheckmarkBold & \color{red}\CheckmarkBold & \color{red}\XSolidBrush & \color{red}\CheckmarkBold & \color{red}\CheckmarkBold \\
\hline
\end{tabular}
\end{center}
\caption{We compare the characteristics and features of our approach with prior methods.  We highlight the unique capabilities of our approach.}
\label{tab:comp}
\end{table*}

In this section, we qualitatively and quantitatively compare the results of our network with prior learning-based methods. We also perform some ablation experiments to validate the effectiveness of our network.

\subsection{Comparisons with Prior Learning Methods}

Many approaches have been proposed to predict cloth deformations using learning-based networks. We have highlighted many recent methods and their attributes in terms of handling different kinds of characters and clothing types in Table~\ref{tab:comp}. Some methods ~\cite{patel2020tailornet, bertiche2021deepsd, bertiche2021pbns,guan2012drape}  are based on the SMPL model, which limits them to only processing SMPL human bodies. ~\cite{santesteban2019learning, santesteban2021self} are also based on the SMPL model. However, it is possible to extend them to remove the dependence on SMPL-based representation. Therefore, we modify these two methods and compare their results with our method in the following sections. ~\cite{holden2019subspace} uses PCA to extract the principal components of the character vertices and cloth vertices to learn the relationship with the next deformation in the subspace. However, this method uses the previous prediction as the input for subsequent predictions and may result in accumulated errors. ~\cite{gundogdu2019garnet, gundogdu2020garnet++} use DQS~\cite{Kavan07} to pre-deform the cloth mesh from the canonical pose to the target pose and then use a learning-based network to predict the residual of the pre-deformed cloth mesh and ground truth. This method only works well on tight-fitting cloth, and its predictions tend to be smooth and may lose wrinkle details. ~\cite{wang2019learning} use MLP to learn the intrinsic features for cloth vertices and character vertices, which results in a model with many redundant parameters. Furthermore, these methods are mostly limited to one or many specific characters or clothing types. In contrast, our network can overcome these limitations and is more general.

\subsection{Qualitative Comparisons}

We have implemented the modified versions of ~\cite{bertiche2021deepsd} and ~\cite{bertiche2021pbns} to process the non-SMPL characters. We replace the SMPL skinning method with a character skinning method, which is based on using skeletons. The modified version of ~\cite{bertiche2021pbns} is an unsupervised method. ~\cite{bertiche2021deepsd} contains the supervised part and the unsupervised part in its network. We have compared our network with the supervised part of ~\cite{bertiche2021deepsd}.

\begin{figure*}[h]
  \centering
  \includegraphics[width=0.95\linewidth]{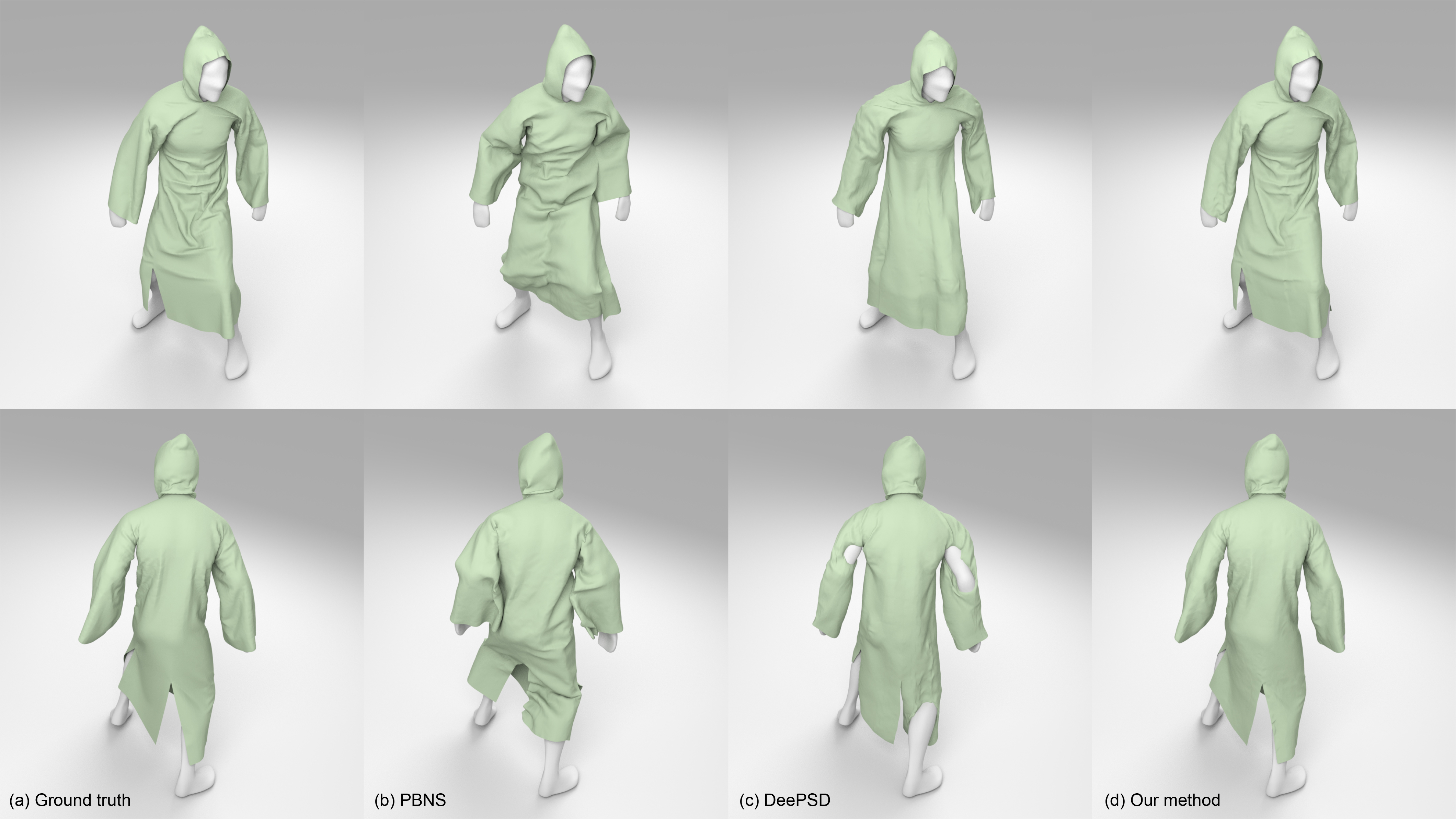}
  \caption{Comparison of results between our network and previous methods. The first column is the ground truth of the deformed cloth. The second and third columns are the results of ~\cite{bertiche2021pbns} and ~\cite{bertiche2021deepsd}. The last column is the result of our method. The top and bottom rows are the front and back views of the deformed predictions.}
  \label{fig:pbns}
\end{figure*}

Fig.~\ref{fig:pbns} shows the comparison between the prediction of PBNS~\cite{bertiche2021pbns}, DeePSD~\cite{bertiche2021deepsd}, and our method. As shown in Fig.~\ref{fig:pbns}, the PBNS method ~\cite{bertiche2021pbns} tends to predict the deformed cloth mesh, which is tightly wrapped on the character and can introduce artifacts in the deformation. The DeePSD method ~\cite{bertiche2021deepsd} tends to predict smooth deformations, resulting in penetrations with the character even after post-processing. This implies that the prediction of DeePSD~\cite{bertiche2021deepsd} is driven less by the transformation matrix of the character. In contrast, the results of our method tend to generate deformations with fine wrinkles. We have also implemented the learning algorithm~\cite{li2021n} and obtained similar results with our method. The prediction results of ~\cite{li2021n} can also generate fine wrinkles. However, ~\cite{li2021n} uses significantly higher memory footprint (about $928.8$MB).

\subsection{Quantitative Comparisons}

We also perform quantitative comparisons between our method and previous methods. We use the following error metrics to evaluate the prediction results of our network and others.
\begin{equation}
\begin{aligned}
\mathcal{E}_{dist} &= \frac{1}{N} \sum_{i=1}^{N} \left\|x_{p}^{i}-x_{g}^{i}\right\|, \\
\mathcal{E}_{norm} &= \frac{1}{N} \sum_{i=1}^{N} \arccos \left(\frac{(n_{p}^{i})^{T}n_{g}^{i}}{\left\|n_{p}^{i}\right\|\left\|n_{g}^{i}\right\|}\right), \\
\end{aligned}
\end{equation}
where $x_{p}^{i}$ is the position of vertex $i$ of the predicted mesh $P$. $x_{g}^{i}$ is the ground truth of vertex $i$. $N$ is the number of vertices of the cloth mesh.
$n_{p}^{i}$ and $n_{g}^{i}$ are the normal vectors of vertex $i$ on the predicted mesh and the ground truth, respectively.

\begin{table}[h]
\begin{center}
\begin{tabular}{|l|c|c|c|}
\hline
Evaluation & PBNS & DeePSD & Our Method \\
\hline\hline
mean $\mathcal{E}_{dist}$(m) & 7.350E-2 & 3.10E-2 & 1.08E-2 \\
std $\mathcal{E}_{dist}$(m) & 1.05E-2 & 6.05E-3 & 2.11E-3 \\
mean $\mathcal{E}_{norm}$($^{\circ}$) & 42.44 & 31.72 & 9.12 \\
std $\mathcal{E}_{norm}$($^{\circ}$) & 3.26 & 3.28 & 1.57 \\
\hline
\end{tabular}
\end{center}
\caption{We compare the mean and standard deviations of mesh errors on test samples based on the ground truth computed from physics-based simualtors.}
\label{tab:error}
\end{table}

\begin{figure}[h]
  \centering
  \includegraphics[width=0.9\linewidth]{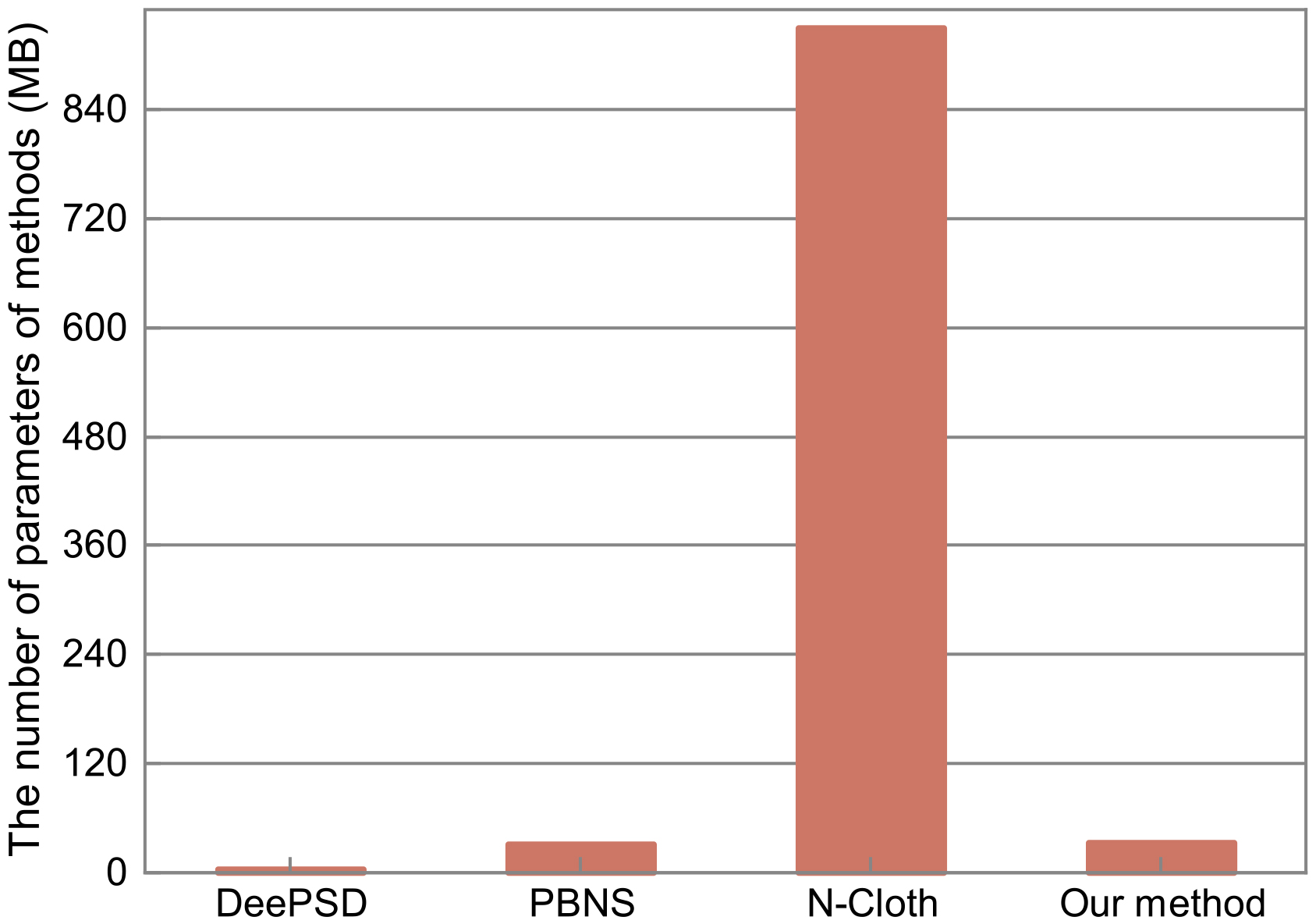}
  \caption{Our approach can is general in terms of handling all skeleton-based models and meshes, but has low memory overhead.
  }
  \label{fig:nparameter}
\end{figure}

The calculated error metrics are shown in Table~\ref{tab:error}. The results generated from our network are more accurate than PBNS~\cite{bertiche2021pbns} and DeePSD~\cite{bertiche2021deepsd}.

We also compare the memory footprint (i.e., number of parameters used) of different networks in Fig.~\ref{fig:nparameter} by measuring the model size. Compared with ~\cite{li2021n}, whose memory footprint is $928.8$MB, the memory footprint of our method is much less ($36.5$MB). The memory footprint of DeePSD  is $3.22$MB,  and PBNS is $30.4$ MB.

\subsection{Ablation Experiments}

\begin{figure*}[h]
  \centering
  \includegraphics[width=\linewidth]{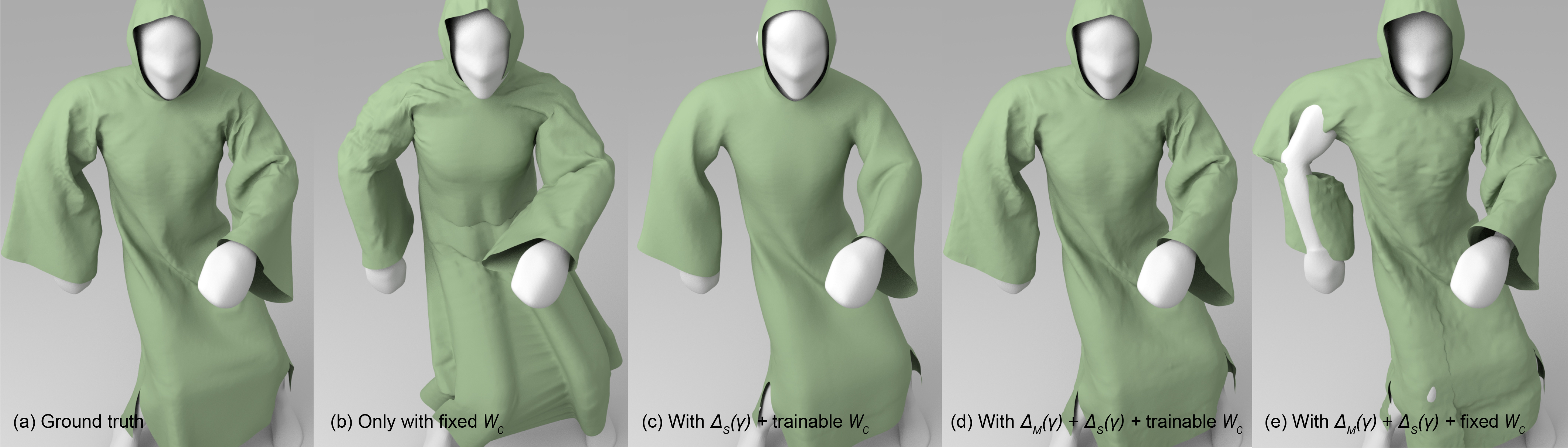}
  \caption{The ablation experiments of our network. We have disabled the mesh-based residual stream, the skeleton-based residual stream, and the trainable cloth weights in our method to show the benefits of each component of our architecture.}
  \label{fig:ablation}
\end{figure*}

To validate the effectiveness of our network architecture, we implement a series of ablation experiments. Fig.\ref{fig:ablation} shows the results of the modified network without some parts of the overall architecture. Fig.~\ref{fig:ablation} (a) is the ground truth of the deformed cloth. Fig.~\ref{fig:ablation} (b) is the cloth skinning deformation only with the fixed initial skinning weight. With the fixed skinning weight, there are artifacts on the skinning deformation, such as legs and belly. Fig.~\ref{fig:ablation} (c) is the result with the skeleton-based residual stream and trainable cloth skinning weight. The deformation in Fig.~\ref{fig:ablation} (c) tends to obtain the coarse residual. Fig.~\ref{fig:ablation} (d) is the result of our full network architecture with the skeleton-based residual stream, the mesh-based residual stream and trainable cloth skinning weight. Compared with the result of Fig.~\ref{fig:ablation} (c), Fig.~\ref{fig:ablation} (d) shows that our mesh-based residual stream can capture the fine details of the final deformation. Fig.~\ref{fig:ablation} (e) is the result of our network without the trainable cloth skinning weight. Without the trainable skinning weight, the skinning result tends to predict more artifacts. There are folds on the legs similar with the result of Fig.~\ref{fig:ablation} (b).


The parameter $m$ for the skeleton-based residual stream and $k$ for mesh-based residual stream also impacts the performance. We have used different values of $m$ and $k$ to train our network. With the increase in the value of $m$, the prediction of our network becomes more accurate. However, the memory footprint also increases, which increases the model size of our network. We show the relevant memory footprint of our network on the scene of Qman dressing robe with $m = 5, 32, 100$ and $k = 64, 128, 256$, respectively in Table.~\ref{tab:msize}. We choose $m = 32$ and $k = 128$ by experiments and find that increasing $m$ and $k$ does not obviously improve the results.


\begin{table}
\begin{center}
\begin{tabular}{|l|c|c|}
\hline
Modified network & Memory footprint (MB)  \\
\hline\hline
$m = 5, k = 64$  & 25.8  \\
$m = 32, k = 128$  & 36.5  \\
$m = 100, k = 256$  & 59.5 \\
\hline
\end{tabular}
\end{center}
\caption{Memory footprint with different $m$ and $k$.}
\label{tab:msize}
\end{table}

\section{Conclusions, Limitations and Future Work}

We present a two-stream skinning-based network to predict cloth deformation from a template cloth in a canonical pose. Our method can process different characters and cloth types retaining the fine details. Since our network is based on the skinning operation, the memory footprint of our method is low. The runtime performance of our network is fast, and we can predict a single cloth deformation in $7$ms on a desktop GPU.

Our approach does have some limitations. Like prior learning-based methods, collision-free predictions are not guaranteed by our network. As part of future work, we would like to overcome the above limitations and extend our work to unsupervised networks~\cite{bertiche2021deepsd} or self-supervised networks~\cite{Santesteban22}. In addition, our method tend to train a specific model for each character due to the difference between human and non-human characters.


\section*{Acknowledgement}

This work is supported in part by the National Natural Science Foundation of China under Grant No.: 61972341, Grant No.: 61972342, Grant No.: 61732015, and the Tencent-Zhejiang University joint laboratory.


{\small
\bibliographystyle{cvm}
\bibliography{cvmbib}
}

\end{document}